\documentclass[runningheads]{llncs}
\usepackage{enumerate}
\usepackage{hyperref}
\usepackage{amssymb, amsmath}
\usepackage{graphics,graphicx,color}
\usepackage{wrapfig}
\usepackage{paralist}
\usepackage[labelsep=period,labelfont=bf,size=small,compatibility=false]{caption}
\usepackage[labelfont=default]{subcaption}
\usepackage{times}
\usepackage{multirow}
\usepackage{bigdelim}
\usepackage{xspace}
\usepackage[absolute]{textpos}
\usepackage{booktabs}

\graphicspath{{pic/}}

\newcommand{\sefe}{\textsc{Sefe}\xspace}
\newcommand{\racsim}{\textsc{RacSim}\xspace}
\newcommand{\racsefe}{\textsc{RacSefe}\xspace}
\newcommand{\MA}{\ensuremath{\mathcal{A}}\xspace}

\newcommand{\MC}{\ensuremath{\mathcal{C}}\xspace}
\newcommand{\MG}{\ensuremath{\mathcal{G}}\xspace}

\newcommand{\MM}{\ensuremath{\mathcal{M}}\xspace}
\newcommand{\MO}{\ensuremath{\mathcal{O}}\xspace}
\newcommand{\MP}{\ensuremath{\mathcal{P}}\xspace}

\newcommand{\MT}{\ensuremath{\mathcal{T}}\xspace}

\newcommand{\MW}{\ensuremath{\mathcal{W}}\xspace}
\newcommand{\MZ}{\ensuremath{\mathcal{Z}}\xspace}
\newcommand{\eps}{\ensuremath{\hat e}}

\DeclareMathOperator{\sgn}{sgn}

\spnewtheorem*{sketchofproof}{Sketch of Proof}{\itshape}{\rmfamily}

\begin{document}
\title{Simultaneous Drawing of Planar Graphs \\ with Right-Angle
  Crossings and Few Bends\thanks{This research was supported by the
    ESF EuroGIGA project GraDR (DFG grant Wo~758/5-1).}}

\author{Michael~A.~Bekos\inst{1} \and Thomas~C.~van~Dijk\inst{2}
  \and Philipp~Kindermann\inst{2} \and Alexander~Wolff\inst{2}}

%
\authorrunning{M.~A.~Bekos, T.~C.~van~Dijk, P.~Kindermann, A.~Wolff}
%
\tocauthor{M.~A.~Bekos, T.~C.~van~Dijk, P.~Kindermann, A.~Wolff}
%
\authorrunning{M.~A.~Bekos, T.~C.~van~Dijk, P.~Kindermann, A.~Wolff}

\institute{%
    Wilhelm-Schickard-Institut f\"ur Informatik, Universit\"at T\"ubingen, Germany\\
    \email{bekos@informatik.uni-tuebingen.de}
    \and
    Lehrstuhl f\"ur Informatik I, Universit\"at W\"urzburg, Germany.\\
    \email{http://www1.informatik.uni-wuerzburg.de/en/staff}
}
\maketitle

\begin{abstract}
  Given two planar graphs that are defined on the same set of
  vertices, a \emph{RAC simultaneous drawing} is a drawing of the two
  graphs where each graph is drawn planar, no two edges overlap, and
  edges of one graph can cross edges of the other graph only at right
  angles.  In the geometric version of the problem, vertices are drawn
  as points and edges as straight-line segments.  It is known,
  however, that even pairs of very simple classes of planar graphs
  (such as wheels and matchings) do not always admit a geometric RAC
  simultaneous drawing.

  In order to enlarge the class of graphs that admit RAC simultaneous
  drawings, we allow edges to have bends.  We prove that any pair of
  planar graphs admits a RAC simultaneous drawing with at most six
  bends per edge.  For more restricted classes of planar graphs (e.g.,
  matchings, paths, cycles, outerplanar graphs, and subhamiltonian
  graphs), we significantly reduce the required number of bends per
  edge.  All our drawings use quadratic area.
\end{abstract}

\section{Introduction}\label{sec:intro}

A simultaneous embedding of two planar graphs embeds each graph in
a planar way---using the same vertex positions for both embeddings.
 Edges of one graph are allowed to intersect edges of the other
graph.  There are two versions of the problem: In the first
version, called \emph{Simultaneous Embedding with Fixed Edges}
(\sefe), edges that occur in both graphs must be embedded
in the same way in both graphs (and hence, cannot be crossed by any
other edge).  In the second version, simply called \emph{Simultaneous
Embedding}, these edges can be drawn differently for each of the
graphs. Both versions of the problem have a geometric variant where
edges must be drawn using straight-line segments.

Simultaneous embedding problems have been investigated extensively
over the last few years, starting with the work of Brass et
al.~\cite{bcdeeiklm-spge-CG07} on simultaneous straight-line
drawing problems.  Bl\"asius et al.~\cite{bkr-sepg-HGDV13} recently
surveyed the area. For example, it is possible to decide in linear
time whether a pair of graphs admits a \sefe or not, if the
common graph is biconnected~\cite{adfpr-tse2g-JDA12} or if it has a fixed
planar embedding~\cite{adfjkpr-tppeg-TALG15}. Furthermore,
\sefe can be decided in polynomial time if each connected
component of the common graph is biconnected or
subcubic~\cite{s-ttpht-JGAA13}, or if it is outerplanar with cutvertices of
degree at most~3~\cite{bkr-seeor-GD13}.

When actually drawing these simultaneous embeddings, a natural
choice is to use straight-line segments. It is NP-hard
to decide whether two planar graphs admit a geometric simultaneous
embedding~\cite{egjpss-sgge-GD08}. This negative results holds
even if one of the input graphs is a matching~\cite{cklmsv-gsegm-JGAA11}.
In fact, only very few graphs can be drawn in this way and some existing
results require exponential area. For instance, there exist a tree
and a path which cannot be drawn simultaneously with straight-line
segments~\cite{agkn-tpgse-12}, and the algorithm for simultaneously
drawing a tree and a matching~\cite{cklmsv-gsegm-JGAA11} does not
provide a polynomial area bound.

A way to overcome the restrictions of straight-line drawings is to
allow edges to bend.  The resulting drawings with polygonal edges are
called \emph{polyline drawings} for short.  Such drawings have been
investigated by Haeupler et al.~\cite{hrl-tspcg-JGAA13}.  They showed
that if the common graph
is biconnected, then a drawing can be found in which one input graph has
no bends at all and in the other input graph the number of bends per edge
is bounded by the number of common vertices.
Erten and Kobourov~\cite{ek-sefb-JGAA05}
showed that three bends per edge and quadratic area suffice for any
pair of planar graphs (without fixed edges), and that one bend per
edge suffices for pairs of trees. Kammer~\cite{k-setbe-SWAT05}
reduced the required number of bends per edge to two for the
general case of planar graphs. Grilli et al.~\cite{ghkr-dwegf-GD14}
proved that every \sefe embedding of two planar graphs admits a
drawing with no bends in the common edges and at most nine bends per
exclusive edge.  For the case that the common graph is biconnected, 
they reduced the number of bends to three per exclusive edge.
Very recently, Frati et al.~\cite{fhk-sefbc-GD15} improved upon these
\sefe results.  They reduced the bend complexity for pairs of planar
graphs to~6 and for pairs of trees to~1.  They also bounded the number
of crossings per edge pair; to~16 and~4, respectively.  In that
respect, Frati et al.\ also improve upon a result of Chan et
al.~\cite{cfglms-mlepg-GD14} who needed 24 crossings per edge pair.
While Chan et al.\ used up to $72n$ bends per exclusive edge, they
managed to keep their drawing on a grid of polynomial size ($O(n^2)
\times O(n^2)$), which is neither the case for Frati et
al.~\cite{fhk-sefbc-GD15} nor for Grilli et
al.~\cite{ghkr-dwegf-GD14}.
In all these results, however, the \emph{crossing angles} can be very
small.  Additionally, the \sefe drawings with $O(1)$ bends per edge
use at least exponential space.

In this paper, we suggest a new approach that overcomes the
aforementioned problems. We insist that crossings occur at right
angles, thereby ``taming'' them. We do this while drawing all
vertices and all bends on an integer grid of size $O(n) \times
O(n)$ for any pair of planar $n$-vertex graphs.  
In a way, our results give a measure for the
geometric complexity of simultaneous embeddability for various
combinations of graph classes, some of which can be combined more easily
(that is, with fewer bends) and some not as easily (that is,
with more bends).

Let~$G_1=(V,E_1)$ and~$G_2=(V,E_2)$ be two planar
graphs defined on the same vertex set and let $n=|V|$.
We say that~$G_1$ and~$G_2$
admit a \emph{RAC simultaneous drawing} (or, \emph{\racsim drawing}
for short) if we can place the vertices on the plane such that:
\begin{enumerate}[(i)]
\item each edge is drawn as a polyline,
\item both graphs are drawn planar,
\item non-common edges are either disjoint or cross each other (one or
  several times) at right angles, and
\item common edges may be represented by the same polyline. 
\end{enumerate}

Note that non-common edges may not overlap. 
$G_1$ and~$G_2$ admit a \emph{RAC simultaneous drawing with fixed
edges} (or, \emph{\racsefe drawing} for short) if they admit a \racsim drawing
with the adjusted condition (iv):
\begin{enumerate}[(i')]
	\setcounter{enumi}{3}
  \item common edges \emph{must} be represented by the same polyline.
\end{enumerate}

\begin{figure}[tb]
  \centering
		\begin{subfigure}[t]{.44\textwidth}
			\centering
			\includegraphics[page=1]{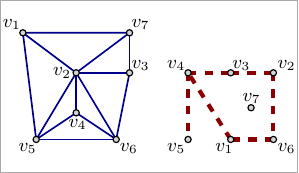}
			\caption{two graphs on the same vertex set}
			\label{fig:definition-1}
		\end{subfigure}
		\begin{subfigure}[t]{.26\textwidth}
			\centering
			\includegraphics[page=2]{definition}
			\caption{a \racsim drawing}
			\label{fig:definition-2}
		\end{subfigure}
		\begin{subfigure}[t]{.28\textwidth}
			\centering
			\includegraphics[page=3]{definition}
			\caption{a \racsefe drawing}
			\label{fig:definition-3}
		\end{subfigure}
		\caption{(a)~Two planar graphs on the same vertex set. 
      (b)~A \racsim drawing of the two 
      graphs with at most one bend per edge. The edge $(v_2,v_6)$
      is drawn differently in the two graphs, while the edges $(v_3,v_4)$ and 
      $(v_2,v_3)$ are both represented by the same polyline in the two graphs. 
      (c)~A \racsefe drawing of the two graphs with up to two bends per edge.}
    \label{fig:definition}
\end{figure}

In Figure~\ref{fig:definition}, we give an example of two planar graphs
on the same vertex set that admit a \racsim drawing with at most one bend
per edge and a \racsefe drawing with at most two bends per edge.

Argyriou et al.~\cite{abks-grsdg-JGAA13} introduced and studied the
geometric version of the \racsim drawing problem. In particular,
they proved that any pair of a cycle and a matching admits a
geometric \racsim drawing on an integer grid of quadratic size,
while there exist a wheel and a cycle that do not
admit a geometric \racsim drawing.  The problem that we study 
in this paper was left as an open problem.

Closely related to the \racsim drawing problem is the problem of
simultaneously drawing a (primal) embedded graph and its dual, so
that the primal-dual edge crossings form right angles.
Brightwell and Scheinermann~\cite{bs-rpg-93} proved that this
is always possible if the input graph is
triconnected planar. Erten and Kobourov~\cite{ek-sepgd-05}
presented an $O(n)$-time algorithm that computes a simultaneous
drawing of a triconnected planar graph and its dual on an integer
grid of size $O(n)\times O(n)$, where $n$ is the total number of
vertices in the graph and its dual; their drawings, however, can
have non-right angle crossings.

\paragraph{Our contribution.}

\begin{table}
  \centering
  \begin{tabular}{@{}l@{\hspace{1ex}$+$\hspace{1ex}}lccl@{}}
    \toprule
    \multicolumn{2}{c}{Graph classes} & Number of bends & Ref. &
    \multicolumn{1}{c}{Model} \\
    \midrule
    planar & planar & $6+6$ & Thm.~\ref{thm:planar} & \racsim\\
    subhamiltonian & subhamiltonian & $4+4$ & Cor.\hfill\ref{cor:twopage} & \racsim\\
    outerplanar & outerplanar & $3+3$ & Thm.~\ref{thm:outerplanar} & \racsim\\
    \midrule
    cycle & cycle & $1+1$ & Thm.~\ref{thm:cycle} & \racsefe\\
    caterpillar & cycle & $1+1$ & Thm.~\ref{thm:cater} & \racsefe\\
    \multicolumn{2}{@{}l}{four matchings} & $1+1+1+1$ &
    Thm.~\ref{thm:fourmatch} & \racsim\\ 
    tree & matching & $1+0$ & Thm.~\ref{thm:treematch} & \racsefe\\
    \midrule
    wheel & matching & $2+0$ & Thm.~\ref{thm:wheelmatch} & \racsefe\\
    outerpath & matching & $2+1$ & Thm.~\ref{thm:outpmatch} & \racsefe\\
    \bottomrule
  \end{tabular}
  \caption{A short summary of our results}
  \label{table:res}
\end{table}

Our main result is that any pair of planar graphs admits a \racsim
drawing with at most six bends per edge. For pairs of subhamiltonian
graphs and pairs of outerplanar graphs, we can reduce the required
number of bends per edge to four and three, respectively; see
Section~\ref{sec:morebends}. (Recall that a subhamiltonian graph is
a subgraph of a planar Hamiltonian graph.) Then, we turn our
attention to pairs of more restricted graph classes where we can
guarantee \racsim and \racsefe drawings with one bend per edge or two bends per
edge; see Sections~\ref{sec:onebend} and~\ref{sec:twobends},
respectively. Table~\ref{table:res} summarizes our results. Note
that all our algorithms run in linear time, with the exception of
the algorithm for an outerpath and a matching (see Theorem~\ref{thm:outpmatch}),
which runs in~$O(n\log n)$ time. The produced drawings fit
on integer grids of quadratic size. The main approach of all our
algorithms is to find linear orders on the vertices of the two
graphs and then to compute the exact coordinates of the vertices of
both graphs based on these orders.
These drawings may
contain non-rectilinear segments (referred to as \emph{slanted}
segments, for short), but all crossings in our drawings appear between
horizontal and vertical edge segments and are therefore at right angle.

\section{\racsim Drawings of General Graphs}
\label{sec:morebends}

In this section, we study general planar graphs and show how to
efficiently construct \racsim drawings in quadratic area, with
few bends per edge. We prove that two planar graphs on a common set
of vertices admit a \racsim drawing with six bends per edge
(Theorem~\ref{thm:planar}). 
We lower the required number of bends per edge to~$4$ for pairs
of subhamiltonian graphs (Corollary~\ref{cor:twopage}), and to~$3$
for pairs of outerplanar graphs (Theorem~\ref{thm:outerplanar}).
Note that the class of
subhamiltonian graphs is equal to the class of 2-page
book-embeddable graphs, and the class of outerplanar graphs is
equal to the class of 1-page book-embeddable
graphs~\cite{bh-btg-JCT79}.

\begin{figure}[ht]
  \centering
  \includegraphics{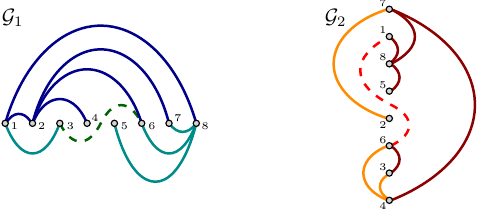}
  \caption{~Drawings of two planar graphs by Kaufmann and 
    Wiese~\cite{kw-evpfb-JGAA02}.}
  \label{fig:planar-1}
\end{figure}  

Central to our approach is an algorithm by Kaufmann and
Wiese~\cite{kw-evpfb-JGAA02} that embeds any planar graph such that
vertices are mapped to points on a horizontal line (the so-called
\emph{spine}) and each edge crosses the spine at most once; see
Figure~\ref{fig:planar-1}. If one replaces each spine
crossing with a dummy vertex, then a \emph{linear order} of the
vertices (both original and dummy) is obtained with the property that
every edge is either completely above or completely below the
spine. In order to determine the exact locations of the vertices of
the two given graphs in our problem, we use the linear
order induced by the first graph to compute the $x$-coordinates and
the linear order induced by the second graph to compute the
corresponding $y$-coordinates. (Note that this
approach has been used for simultaneous drawing problems
before~\cite{ek-sefb-JGAA05}.) Then, we draw the edges of both
graphs, so that all edges-crossings
\begin{enumerate}[(i)]
\setlength{\itemsep}{0mm}
\item are restricted between vertical and horizontal edge-segments,
that is, slanted edge-segments are crossing-free, and
\item appear in the interior of the smallest axis-aligned rectangle
containing all vertices.
\end{enumerate}

\begin{theorem}\label{thm:planar}
  Two planar graphs on a common set of $n$ vertices admit a \racsim drawing
  on an integer grid of size $(14n-26) \times (14n-26)$
  with six bends per edge.
  The drawing can be computed in $O(n)$ time.
\end{theorem}
\begin{proof}
Let~$\MG_1=(V,E_1)$ and~$\MG_2=(V,E_2)$ be planar graphs.
For $m\in\{1,2\}$, let~$\xi_m$ be an embedding of~$\MG_m$ according to
the algorithm of Kaufmann and Wiese~\cite{kw-evpfb-JGAA02}; see
Figure~\ref{fig:planar-1}. 
As explained before, we subdivide all edges of $\MG_m$ that cross the
spine in $\xi_m$ by introducing a dummy vertex
at the point where it crosses the spine.
Let~$\MG_m'=(V_m',E_m')=(V \cup V_m, A_m \cup B_m)$ be the
resulting graph, where~$V_m$ is the set of dummy vertices
in~$\MG_m'$, $A_m$ is the set of edges that are drawn completely
above the spine and $B_m$ is the set of edges that are drawn
completely below the spine. Let $\xi_m'$ be the embedding
of~$\MG_m'$.

Now we show how to determine the $x$-coordinates of
the vertices in~$V_1'$ by assigning the vertices to the columns
of a grid; the $y$-coordinates of the vertices
in~$V_2'$ are determined analogously. Let~$n_1'$ be the number of
vertices in~$V_1'$ and let $v_1 \rightarrow v_2 \rightarrow \dots
\rightarrow v_{n_1'}$ be the linear order of the vertices of~$V_1'$
along the spine in~$\xi_1'$.  We start by placing~$v_1$. 
Between any two consecutive vertices~$v_i$
and~$v_{i+1}$, we reserve several columns for the bends of the
edges incident to~$v_i$ and~$v_{i+1}$; see Figure~\ref{fig:planar-2}.
The columns are used (in the given order) for the following purposes:
\begin{enumerate}[(i)]
\item \label{col:1} for the first bend on all edges in~$A_2$ leaving~$v_i$,
\item \label{col:2} for the first vertical segment of each
  edge~$(v_i,v_j)\in E_1'$ with~$j>i$, 
\item \label{col:3} for the last vertical segment of each
  edge~$(v_k,v_{i+1})\in E_1'$ with~$k \le i$, and 
\item \label{col:4} for the last bend on all edges in~$B_2$ entering~$v_{i+1}$.
\end{enumerate}
Note that we can save some columns reserved for
(\ref*{col:2}) and (\ref*{col:3}) because an edge in~$A_1$ and an
edge in~$B_1$ can use the same column for their bend.

\begin{figure}[tb]
  \centering
  \includegraphics{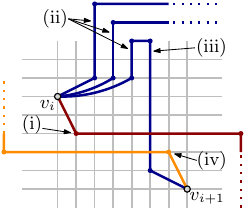}
  \caption{Reserving additional columns between~$v_i$ and~$v_{i+1}$.}
  \label{fig:planar-2}
\end{figure}

With this procedure we fully specify the $x$-coordinates of
the vertices in $V_1'$; the $y$-coordinates of the vertices
in~$V_2'$ are determined analogously. Let~$R$ be the smallest
axis-aligned rectangle that contains all vertices of the common
vertex set of~$\MG_1'$ and~$\MG_2'$ (the gray
rectangle in Figure~\ref{fig:planar-3}). Note that the $y$-coordinates
of the dummy vertices of~$V_1'$ and the $x$-coordinates of the
dummy vertices of~$V_2'$ have not been determined by
the algorithm so far. They can be set arbitrarily, as long as the
corresponding vertices are inside~$R$. 

We proceed to describe how to draw the edges of graph $\MG_1'$ with
at most four bends per edge such that all edge segments of~$\MG_1'$
in~$R$ are either vertical or of $y$-length exactly~1; see
Figure~\ref{fig:planar-4}. The edges of graph $\MG_2$ are drawn
analogously (rotated by~$90^\circ$). First, we draw the edges~$(v_i,v_j)\in A_1$ with~$i<j$
in a nested order: When we draw the edge~$(v_i,v_j)$, there is no
edge~$(v_k,v_l)\in A_1$ with~$k\le i$ and~$l\ge j$ that has not
already been drawn. Recall that the first column to the right and
the first column to the left of every vertex is reserved for the
edges in~$E_2$, hence we assume that they are already used.
We draw~$(v_i,v_j)$ with at most four bends as follows. We start
with a slanted segment incident to $v_i$ that has its other
endpoint in the row above~$v_i$ and in the first unused column that
does not lie to the left of~$v_i$. We follow with an upward vertical
segment that leaves~$R$. We add a horizontal segment
above~$R$; the row of this segment is determined by the nesting in the
book embedding of $\MG'_1$.
In the last unused column that does not lie to the right
of~$v_j$, we add a vertical segment that ends one row above~$v_j$.
We finish the edge with a slanted segment that has its endpoint
in~$v_j$. We draw the edges in~$B_1$ symmetrically, with the
horizontal segment below~$R$.

\begin{figure}[tb]
    \begin{subfigure}[t]{.47\textwidth}
      \centering
      \includegraphics[page=1]{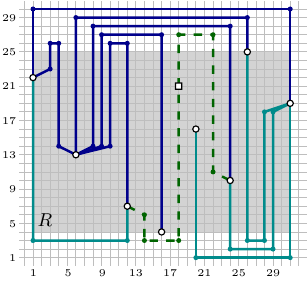}
      \caption{$\MG_1$}
      \label{fig:planar-3}
    \end{subfigure}
    \hfill
    \begin{subfigure}[t]{.47\textwidth}
      \centering
      \includegraphics[page=2]{planar}
      \caption{$\MG_1$ and $\MG_2$}
      \label{fig:planar-4}
    \end{subfigure}

    \caption{\racsim drawings of the graphs~$\MG_1$ and $\MG_2$ (see
      Figure~\ref{fig:planar-1}) with at most six bends per edge,
      generated by our algorithm.  The edges that cross the spine are
      drawn dashed; the dummy vertices on these edges are drawn as
      squares.}
    \label{fig:planar}
\end{figure}

Note that this algorithm always uses the top and the bottom port
of a vertex~$v$ if there is at least one edge incident to~$v$
in~$A_1$ and~$B_1$ respectively.
The edges incident to a dummy vertex use precisely the top and the bottom port:
Dummy vertices have exactly one incident edge in~$A_1$ and one in~$B_1$.
We create a drawing of~$\MG_1$ and~$\MG_2$ with at most~6
bends per edge by bypassing (that is, smoothing out) the dummy 
vertices. This does not change the drawing.

By construction, all edge segments of~$E_1$ inside~$R$ are either
vertical segments or slanted segments of $y$-length~1.
Symmetrically, all segments of~$E_2$ inside~$R$ are either
horizontal segments or slanted segments of $x$-length~1. Thus, the
slanted segments cannot intersect. All crossings
inside~$R$ occur between a horizontal and a vertical segment, and
thus form right angles. Also, there are no segments in~$E_1$ that
lie to the left or to the right of~$R$, and there are no segments
in~$E_2$ that lie above or below~$R$. Hence, there are no crossings
outside~$R$, which guarantees that the constructed drawing
of~$\MG_1$ and~$\MG_2$ is a \racsim drawing.

We now count the columns used by the drawing. For the leftmost and
the rightmost vertex, we reserve one additional column for its
incident edges in~$E_2$; for the remaining vertices, we reserve two
such columns. For each edge in~$E_1$, we use up to three columns:
one for each endpoint of the slanted segment at each vertex and one
for the vertical segment that crosses the spine, if it exists. Recall
that at least one edge per vertex does not need a slanted segment.
For each edge in~$E_2$, we need at most one column for the vertical
segment to the side of~$R$. Since there are at most~$3n-6$ edges,
we need at most \[3n-2+3 \cdot (3n-6)-n+3n-6=14n-26\] columns. By
symmetry we need the same number of rows.

The algorithm of Kaufmann and Wiese computes a drawing with at most~3
bends per edge in~$O(n)$ time. We can compute the nested order of the edges
in linear time from the embedding, as the circular order of the edges around a 
vertex gives a hierarchical order on the edges that describes the nested order
of the edges. Thus, our algorithm also runs in~$O(n)$ total time.
\end{proof}

We can improve the result of Theorem~\ref{thm:planar} for
subhamiltonian graphs, which admit 2-page book embeddings in which
no edges cross the spine~\cite{bh-btg-JCT79}. Since those edges are
the only ones that need six bends, the number of bends per edge
is reduced to four. The number of columns and rows are also
reduced by one per edge. This is summarized in the following
corollary. Note, however, that it is NP-hard to decide whether a
given planar graph is subhamiltonian, even for maximal planar
graphs~\cite{a-tcothcpfmpg-PU82}. A 2-page book embedding can be
found in linear time, if the ordering of the vertices along the
spine is given~\cite{hs-rnapk-BMB99}. Several classes of graphs
are known to be
(sub)hamiltonian, for example 4-connected planar
graphs~\cite{nc-c10hc-PG08}, planar graphs without separating
triangles~\cite{ko-etw-AML07}, Halin
graphs~\cite{cnp-hgtsp-MP83}, planar graphs with maximum
degree~3~or~4~\cite{bgr-tpbe4-STACS14,h-afegib-UNC85}.

\begin{corollary}\label{cor:twopage}
  Two subhamiltonian graphs on a common set of $n$ vertices
  admit a \racsim drawing on an integer grid of size $(11n-32) \times (11n-32)$
  with four bends per edge. The algorithm runs in~$O(n)$ time if the
  subhamiltonian cycles of both graphs are given.
\end{corollary}

We use even fewer bends for a \racsim drawing of two outerplanar graphs.
This is based on a decomposition of each of the graphs into two forests.
The following lemma shows that we can do this in linear time.

\begin{lemma}\label{lem:outerforests}
  Every outerplanar graph can be decomposed into two forests. This decomposition
  can be computed in linear time.
\end{lemma}
\begin{proof}
  It follows by Nash-Williams' formula~\cite{nw-dfgf-JLMS64}
  that every outerplanar graph has arboricity~2, that is, it can be
  decomposed into two forests.
  To prove the linear running time, we first assume biconnectivity
  and augment the input graph to a maximal outerplanar graph. Now, if we
  add a new vertex that is incident to all vertices of the graph,
  the result is a maximal planar graph which can be decomposed
  into three trees~\cite{f-ggaa-04}, so that one of them is a star
  incident to the newly added vertex. Hence, the removal of this
  vertex yields a decomposition into two trees. The desired
  decomposition into two forests follows from the removal of the
  edges added to augment the graph to maximal outerplanar. For an
  outerplanar graph that is not biconnected, we have to compute
  the aforementioned decomposition for each of its biconnected
  components individually.
  Since the biconnected components of a graph form a tree (the
  so-called \emph{BC-tree}), the structure of the overall
  decomposition is not affected; it still consists of two forests.
  The overall decomposition can be computed in linear time since
  both the decomposition of a maximal planar graph into three trees and
  the computation of the
  biconnected components of the outerplanar input graph can be found
  in linear time.
\end{proof}

With this lemma, we can further improve the required number of bends per edge to
three for outerplanar graphs. (Recall that all outerplanar graphs are
1-page book embeddable.) We will use the order of the vertices on
the spine of a 1-page book embedding to compute a 2-page book
embedding in which every edge uses a rectilinear port at one of
its endpoints, enabling us to omit one of its bends.

\begin{theorem}\label{thm:outerplanar}
  Two outerplanar graphs on a common set of $n$ vertices admit a \racsim drawing
  on an integer grid of size $(7n-10) \times (7n-10)$
  with three bends per edge.
  The drawing can be computed in $O(n)$ time.
\end{theorem}
\begin{proof}
Let~$\MO_1=(V,E_1)$ and~$\MO_2=(V,E_2)$ be the given outerplanar
graphs. We will embed $\MO_1$ and $\MO_2$ on
two pages with one forest per page.

To do so, we first create 1-page book embeddings for~$\MO_1$
and for~$\MO_2$ using the linear time algorithm of
Heath~\cite{h-eogsb-SIAM87}. This gives the orders of the
vertices of both graphs along the spine. It follows by
Corollary~\ref{cor:twopage} that, by using the algorithm described
in the proof of Theorem~\ref{thm:planar}, we can create a \racsim
drawing of $\MO_1$ and $\MO_2$ with at most four bends per edge. We
will now show how to adjust the algorithm to reduce the number of
bends by one.

We decompose~$\MO_1$ into two forests~$A_1$ and~$B_1$ according to
Lemma~\ref{lem:outerforests}.
We will draw the edges of~$A_1$ above the spine and the edges~$B_1$
below the spine. By rooting the trees in~$A_1$ in arbitrary
vertices, we can direct each edge such that every vertex has
exactly one incoming edge. Recall that, in the drawing produced in
Theorem~\ref{thm:planar}, one edge per vertex can use its top port.
We adjust the algorithm such that every directed edge~$(v,w)$
enters vertex~$w$ from its top port. To do so, we draw the edge
as follows. We start with a slanted segment of $y$-length~1. We
follow with a vertical segment to the top,
a horizontal segment that ends directly above~$w$, and finish
the edge with a vertical segment that enters~$w$ in the top port.
We use the same approach for the edges in~$B_1$, using the bottom
port and treat the second outerplanar graph~$\MO_2$ analogously.

Since every port of a vertex is only used once, the drawing has no
overlaps. We now analyze the number of columns used. For every
vertex except for the leftmost and rightmost, two
additional columns are reserved for the edges in~$E_2$; for the remaining two
vertices, we reserve a single additional column. The edges
in~$E_1$ now only need one column for the bend of the single
slanted segment. For every edge in~$E_2$, we need up to one column
for the vertical segment to the side of~$R$. Since there are at
most~$2n-4$ edges, our drawing needs~$3n-2+2n-4+2n-4=7n-10$
columns. Analogously, we can show that the algorithm needs~$7n-10$
rows. Since the decomposition can be computed in~$O(n)$ time, our algorithm
also requires~$O(n)$ time.
\end{proof}

\section{\racsim and \racsefe Drawings with One Bend per Edge}
\label{sec:onebend}

In this section, we study simple classes of planar graphs and show
how to efficiently construct \racsim and/or \racsefe drawings with
one bend per edge in quadratic area. In particular, we prove that two
cycles (four matchings, resp.) on a common set of $n$ vertices admit
a \racsefe (\racsim, resp.) drawing on an integer grid of size $2n
\times 2n$; see~Theorem~\ref{thm:cycle}
(Theorem~\ref{thm:fourmatch}, resp.).
If the input to our problem is a caterpillar and a cycle, then we
can compute a \racsefe drawing with one bend per edge on an integer
grid of size $(2n-1) \times 2n$; see Theorem~\ref{thm:cater}. For a
tree and a cycle, we can construct a \racsefe drawing with one bend
per tree edge and no bends in the matching edges on an integer grid
of size $n \times (n-1)$; see Theorem~\ref{thm:treematch}.

In the next proof and in a few more places throughout this paper, we
use the following common notation.  For any real~$x$, let the
\emph{sign} of~$x$, $\sgn(x)$, be~0 if $x=0$, $1$ if $x>0$, and $-1$
if $x<0$.

\begin{lemma}\label{lem:path}
    Two paths on a common set of $n$ vertices admit a \racsefe drawing
    on an integer grid of size $2n \times 2n$
    with one bend per edge. The drawing can be computed in
    $O(n)$ time.
\end{lemma}
\begin{proof}
Let~$\MP_1=(V,E_1)$ and~$\MP_2=(V,E_2)$ be the given paths.
To keep the description simple, we first assume that
$\MP_1$ and $\MP_2$ do not share edges. Following standard practices
from the literature~(see for example Brass et
al.~\cite{bcdeeiklm-spge-CG07}), we draw~$\MP_1$ $x$-monotone
and~$\MP_2$ $y$-monotone. This ensures that the drawing of the paths
will individually be planar. We will now describe how to compute the
exact coordinates of the vertices and how to draw the edges of
$\MP_1$ and $\MP_2$, such that all crossings are at right angles
and, more importantly, that no edge segments overlap.

For~$m\in\{1,2\}$ and any vertex~$v\in V$, let~$\pi_m(v)$ be the position
of~$v$ in~$\MP_m$. Then,~$v$ is drawn at the
point~$p(v)=(2\pi_1(v)-1,2\pi_2(v)-1)$; see Figure~\ref{fig:paths+cycles}.
It remains to determine, for each edge $e=(v,v')$, where to place its
bend.  
First, assume that $e \in E_1$ and that~$e$ is directed from its left
endpoint, say $v$, to its right endpoint, say $v'$. Then we place
the bend to the left of $v'$, and one row above or below it depending
on which direction the edge is coming from. To be exact, we place it
at $p(v')-(2,\sgn(y(v')-y(v)))$. Second, assume that $e \in E_2$
and~$e$ is directed from its bottom endpoint, say $v$, to its top
endpoint, say $v'$. Then, we place the bend at~$p(v')-(\sgn(x(v')-x(v)),2)$.

\begin{figure}[tb]
  \centering
  \includegraphics{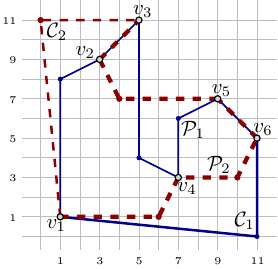}
  \caption{\racsefe drawings with one bend per edge: two paths~$\MP_1$
    (thin solid) and~$\MP_2$ (bold dashed) and two cycles $\MC_1 =
    \MP_1 + (v_1,v_6)$ and $\MC_2 = \MP_2 + (v_1,v_3)$}
  \label{fig:paths+cycles}
\end{figure}

The area required by the drawing is $(2n-1) \times
(2n-1)$. An edge of $\MP_1$ leaves its left endpoint vertically
and enters its right endpoint with a slanted segment of $x$-length~1
and $y$-length~2. Similarly, an edge of $\MP_2$ leaves its bottom
endpoint horizontally and enters its top endpoint with a slanted
segment of $x$-length~2 and $y$-length~1. Hence, the slanted
segments cannot be involved in crossings or overlaps. Since~$\MP_1$
and~$\MP_2$ are $x$- and $y$-monotone, respectively, it follows
that all crossings must involve a vertical edge segment of~$\MP_1$
and a horizontal edge segment of $\MP_2$, which is at right angle.
The runtime is clearly linear.

Finally, we show that our algorithm supports the \sefe model.  
Assume that $\MP_1=(V,E_1)$ and~$\MP_2=(V,E_2)$ share edges,
and let~$e=(v,v')$ be an edge that belongs to both input paths.
Our algorithm automatically places~$v$ and~$v'$ at consecutive $x$- and 
$y$-coordinates and draws $e$ as a diagonal of $x$- and $y$-length~2
in both graphs, which cannot be involved in a crossing. 
\end{proof}

We say that an edge uses the bottom/left/right/top \emph{port} of a
vertex if it enters the vertex from the bottom/left/right/top.

\begin{theorem}\label{thm:cycle}
    Two cycles on a common set of $n$ vertices admit a \racsefe drawing
    on an integer grid of size $2n \times 2n$
    with at most one bend per edge.
    The drawing can be computed in $O(n)$ time.
\end{theorem}
\begin{proof}
Let~$\MC_1=(V,E_1)$ and~$\MC_2=(V,E_2)$ be the given cycles and
assume first that $\MC_1$ and $\MC_2$ do not share edges. Let $v \in
V$ be an arbitrary vertex. We temporarily delete one edge~$(v,w_1) \in
E_1$ from~$\MC_1$ and one edge~$(v,w_2) \in E_2$  
from~$\MC_2$ (refer to the edges $(v_1,v_3)$ and $(v_1,v_6)$ in
Figure~\ref{fig:paths+cycles}). This results into two
paths~$\MP_1=\langle v,\ldots,w_1\rangle$ and~$\MP_2=\langle
v,\ldots, w_2\rangle$. We use the algorithm of
Lemma~\ref{lem:path} to construct a \racsim drawing of~$\MP_1$
and~$\MP_2$ on an integer grid of size $(2n-1) \times (2n-1)$.
Since~$v$ is the first vertex in both paths, it is placed at the
bottom-left corner of the bounding box containing the drawing.
Since~$w_1$ and~$w_2$ are the last vertices in~$\MP_1$ and~$\MP_2$,
respectively,~$w_1$ is placed on the right side, and~$w_2$ on the
top of the bounding box. By construction, the bottom port
of~$w_1$ and the left port of~$w_2$ are both unoccupied. Hence, the
edges $(v,w_1)$ and $(v,w_2)$ that form $\MC_1$ and~$\MC_2$ can be
drawn with a single bend at points $(2n-1,0)$ and $(0,2n-1)$
respectively; see Figure~\ref{fig:paths+cycles}. Since both edges are
completely outside of the bounding box containing the drawing, neither
is involved in crossings. The total area of
the drawing gets larger by a single unit in each dimension.
The runtime bound is unaffected.

It remains to consider the case that~$\MC_1$ and~$\MC_2$
share edges. Since $\MP_1$ and $\MP_2$ already support the \sefe
model, it suffices to consider
the case that a closing edge~$(v,w_1)$ or~$(v,w_2)$ is also contained in the 
other graph. Since the closing edges are drawn planar, we can simply use their
drawing in both graphs and remove the corresponding edge from the path. Thus,
the drawing supports the \sefe model. 
\end{proof}

\begin{theorem}\label{thm:cater}
    A caterpillar and a cycle on a common set of $n$ vertices admit a \racsefe drawing
    on an integer grid of size $(2n-1) \times 2n$
    with one bend per edge.
    The drawing can be computed in $O(n)$ time.
\end{theorem}
\begin{proof}
Let $\MA=(V,E_\MA)$ be the given caterpillar and $\MC=(V,E_\MC)$ the
given cycle. Similar to the previous proofs, we will first prove
that $\MA$ and $\MC$ admit a \racsim drawing, assuming that they do not
share edges. We postpone the case where $\MA$ and $\MC$ share edges
for later. A caterpillar can be decomposed into a path, called its
\emph{spine}, and a set of leaves connected to the path, called its
\emph{legs}. Let $v_1,v_2,\ldots,v_n$ be the vertex set of~$\MA$
ordered as follows (see Figure~\ref{fig:catercycle}): Starting from
an endpoint of the spine of~$\MA$, we traverse the caterpillar such
that we visit all legs incident to a spine vertex before moving on
to the next spine vertex. This order defines the $x$-order of the
vertices in the output drawing.

As in the proof of Theorem~\ref{thm:cycle}, we temporarily delete
an edge of $\MC$ incident to~$v_1$ (see the thin dashed edge in
Figure~\ref{fig:catercycle}) and obtain a path which we denote by
$\MP=(V,E_\MP)$.  For any vertex $v_i \in V$, let $\pi(v_i)$ be the
position of $v_i$ in~\MP.  The map~$\pi$ determines the $y$-order of
the vertices in our drawing. For $i\in\{1,2,\ldots,n\}$, we draw
vertex~$v_i$ at point~$p(v_i)=(2i-1,2\pi(v_i)-1)$. It remains to
determine, for each edge $e=(v,v')$, where to place its bend.
First, assume that $e \in E_\MP$ and that~$e$ is directed from its
bottom endpoint, say $v$, to its top endpoint, say $v'$ (see the
bold dashed edges in Figure~\ref{fig:catercycle}).
Then, we place the bend at $p(v)+(\sgn(x(v')-x(v)),2)$.
Second, assume that $e \in E_\MA$ and~$e$ is directed from its left
endpoint, say $v$, to its right endpoint, say $v'$ (see the solid
edges in Figure~\ref{fig:catercycle}). Then, we place the bend at
$(x(v'),y(v)+\sgn(y(v')-y(v))$.

\begin{figure}[tb]
	\centering
	\includegraphics{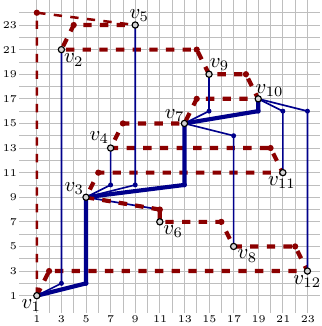}
	\caption{A \racsefe drawing of a caterpillar (solid; its spine is drawn 
		bold) and a cycle (dashed)}
	\label{fig:catercycle}
\end{figure}

The approach described above ensures that $\MP$ is drawn
$y$-monotone, hence planar. The spine of $\MA$ is drawn
$x$-monotone. The legs of a spine vertex of $\MA$ are drawn to the
right of their parent spine vertex and to the left of the next
vertex along the spine. Hence, $\MA$ is drawn planar as well. The
slanted segments of $\MA$ have $y$-length~1, while the slanted
segments of $\MP$ have $x$-length~1. Thus, they cannot be
involved in crossings, which implies that all crossings form right
angles.

It remains to draw the edge~$e$ in $E_\MC \setminus E_\MP$.  Recall
that $e$ is incident to~$v_1$, which lies at the bottom-left corner
of the bounding box containing our drawing. Let $v_j$ be the other
endpoint of~$e$. Since $\pi(v_j)=n$, vertex~$v_j$ lies at the top
of the bounding box. As the top port of~$v_1$ is not used,
we can draw the first segment of~$e$ vertical, bending at~$(1,2n)$;
see the thin dashed edge in Figure~\ref{fig:catercycle}.

To complete the proof of this theorem, it remains to show
that our algorithm supports the \sefe model.  Suppose that there is an
edge~$e=(v,v')$ that belongs to both~\MA and~\MC.  Our algorithm
places~$v$ and~$v'$ at consecutive $y$-coordinates. Thus, the
drawing of~$e$ in~\MA consists of a slanted and a vertical segment
of $y$-length~1 each.  This drawing of~$e$ in~\MA is not crossed by
any edge of~$E_\MC$, so~$e$ can be drawn in the same way for both
graphs.

Clearly, the area of the drawing is $(2n-1) \times 2n$, and the
runtime is linear.
\end{proof}

\begin{theorem}\label{thm:fourmatch}
    Four matchings on a common set of $n$ vertices admit a \racsim drawing
    on an integer grid of size $2n \times 2n$
    with at most one bend per edge.
    The drawing can be computed in $O(n)$ time.
\end{theorem}
\begin{proof}
Let $\MM_1=(V,E_1)$, $\MM_2=(V,E_2)$, $\MM_3=(V,E_3)$
and~$\MM_4=(V,E_4)$ be the given matchings. Without loss of
generality, we assume that all matchings are perfect; otherwise, we
augment them to perfect matchings. Let~$\MM_{1,2}=(V,E_1 \cup E_2)$
and~$\MM_{3,4}=(V,E_3 \cup E_4)$. Since~$\MM_1$ and~$\MM_2$ are
defined on the same vertex set,~$\MM_{1,2}$ is a $2$-regular graph.
Thus, each connected component of~$\MM_{1,2}$ corresponds to a
cycle of even length which alternates between edges of~$\MM_1$
and~$\MM_2$; see Figure~\ref{fig:fourmatchings}. The same holds
for~$\MM_{3,4}$. We will determine the $x$-coordinates of the
vertices from~$\MM_{1,2}$, and the $y$-coordinates
from~$\MM_{3,4}$.

\begin{figure}[tb]
	\centering
	\includegraphics{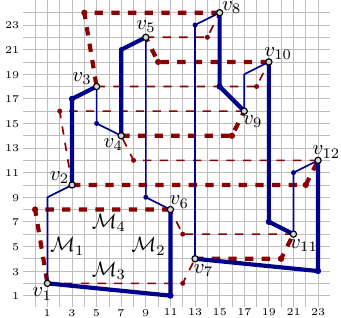}
	\caption{A \racsim drawing of four matchings: $\MM_1$~(solid-plain), 
		$\MM_2$~(solid-bold), $\MM_3$~(dashed-plain), and $\MM_4$ (dashed-bold)}
	\label{fig:fourmatchings}
\end{figure}

We start by choosing an arbitrary vertex~$v \in V$. Let~$\MC$ be
the cycle of~$\MM_{1,2}$ containing vertex~$v$. We determine the
$x$-coordinates of the vertices of~$\MC$ by traversing it in
some direction, starting from vertex $v$. For each vertex~$u$
in~$\MC$, let~$\pi_1(u)$ be the discovery time of~$u$ according to
this traversal, with~$\pi_1(v)=0$. We set~$x(u)=2\pi_1(u)+1$.
After doing this, we determine the $y$-coordinates of all
vertices that lie on cycles in~$\MM_{3,4}$ that contain at least one
vertex of~$\MC$ (that is, cycles in~$\MM_{3,4}$ that contain
a vertex for which the $x$-coordinate has been determined).
Call these cycles~$\MC_1, \ldots, \MC_k$, ordered as follows.
For $i\in\{1,\ldots, k\}$, let~$a_i$ be the \emph{anchor} of~$\MC_i$,
that is, the vertex with the smallest determined $x$-coordinate of
all vertices in~$\MC_i$. Then,~$x(a_1)<\ldots<x(a_k)$. In what
follows, we start with the first cycle~$\MC_1$ of the computed
order and determine the $y$-coordinates of its vertices. To do so,
we traverse $\MC_1$ in some direction, starting from its anchor
vertex~$a_1$. For each vertex~$u$ in~$\MC_1$, let~$\pi_2(u)$ be the
discovery time of~$u$ according to this traversal,
with~$\pi_2(a_1)=0$. Then, we set~$y(u)=2\pi_2(u)+1$. We proceed
analogously with the remaining cycles~$\MC_i$, $i=2,\ldots,k$,
setting~$\pi_2(a_i)=\max_{u\in\MC_{i-1}}\pi_2(u)+1$.

After this step, there are no vertices for which only the
$x$-coordinate has been determined.
However, there might exist vertices where only the $y$-coordinate
has been determined.
If this is the case, we repeat the aforementioned
procedure to determine the $x$-coordinates of the vertices of all
cycles of~$\MM_{1,2} \setminus \MC$ that have at least one vertex
with a determined $y$-coordinate, but without determined
$x$-coordinates. If there are no vertices with only one determined
coordinate left, either all coordinates are determined, or we
restart this procedure with another arbitrary vertex that has no
determined coordinates. Thus, our algorithm guarantees that the
$x$- and $y$-coordinate of all vertices are eventually determined.

Note that for each cycle in~$\MM_{1,2}$ there is exactly one
edge~$e=(v,v')$ with~$\pi_1(v')>\pi_1(v)+1$.
We call this the \emph{closing edge}.
Analogously, for each cycle
in~$\MM_{3,4}$, there is exactly one closing edge~$e=(u,u')$
with~$\pi_2(u')>\pi_2(u)+1$.

Finally we determine, for each edge~$e=(v,v')$, where to place its
bend.  
First, assume that~$e\in E_1\cup E_2$ and that~$e$ is directed from its
left endpoint, say~$v$, to its right endpoint, say~$v'$. If~$e$ is
not a closing edge, we place the bend
at~$(x(v')-2,y(v')-\sgn(y(v')-y(v))$. Otherwise, we place the bend
at~$(x(v'),y(v)-1)$. Second, assume that~$e\in E_3\cup E_4$ and~$e$
is directed from its bottom endpoint, say $v$, to its top endpoint,
say $v'$. If~$e$ is not a closing edge, we place the bend
at~$(x(v')-\sgn(x(v')-x(v)),x(v')-2)$. Otherwise, we place the bend
at~$(x(v)-1,y(v'))$; see Figure~\ref{fig:fourmatchings}.

Our choice of coordinates guarantees that the $x$-coordinates of
the cycles of~$\MM_{1,2}$ and the $y$-coordinates of the cycles
of~$\MM_{3,4}$ form disjoint intervals. Thus,
the area below a cycle of~$\MM_{1,2}$ and the area to the left of a
cycle of~$\MM_{3,4}$ are free from vertices. Hence, the slanted
segments of the closing edges cannot have a crossing that violates
the RAC restriction. The total area required by the drawings
is~$2n\times 2n$. The running time is linear.
\end{proof}

\begin{figure}[tb]
  \centering
  \includegraphics{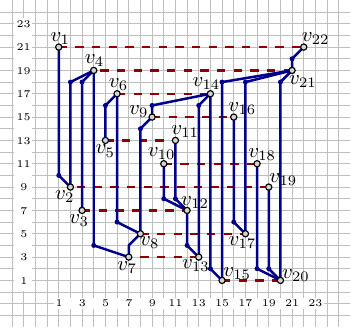}
  \caption{A \racsefe drawing of a tree~(solid; rooted in~$r$) and a
    matching (dashed).  For each of the three subtrees hanging off
    vertex~$s$ we used a different color; the subtrees occupy disjoint
    $x$-intervals (marked below the drawing).}
  \label{fig:treematching}
\end{figure}

We now detail how to compute a \racsefe drawing of a rooted tree
and a matching with one bend per tree edge and no bends on
matching edges.  
Figure~\ref{fig:treematching} shows an example output of our algorithm.
Our layout algorithm is inspired by an algorithm for drawing a
geometric simultaneous embedding of a tree and a matching by 
Cabello et al.~\cite{cklmsv-gsegm-JGAA11}, which in turn goes back
to an algorithm of Di Giacomo et al.~\cite{ddkls-mdpg-JGAA09}.
Cabello et al.\ draw edges straight (that is, no bends), but their
crossing angles are not necessarily right.
We recall some of their notation that we will then use for our purposes.

Cabello et al.\ draw the edges of the matching horizontally.  They 
partition the matching into a \emph{top group} and a \emph{bottom group}. 
Within the top group, they place the edges from top to bottom; vice
versa within the bottom group. Once a matching edge is assigned to one of 
the two groups, the edge and its endpoints are called \emph{placed}.

The assignment of the edges to the groups is iterative.
In each step, the placed vertices induce an edge partition of the tree into 
connected components (subtrees); each component up to and including
the placed vertices is called a \emph{rope}.
If a rope with three placed vertices exists, the
vertex that lies on all three paths between these placed vertices is called a
\emph{splitter}; see Figure~\ref{fig:splitter-def}. Cabello et al.\ showed 
that placing a vertex creates at most one splitter and that placing a
splitter does not create a new one. 

Now, we are ready to present our algorithm in detail.

\begin{theorem}\label{thm:treematch}
  A tree and a matching on a common set of $n$ vertices admit a \racsefe drawing
  on an integer grid of size~$n \times (n-1)$
  with one bend per tree edge and no bends on matching edges.
  The drawing can be computed in $O(n)$ time.
\end{theorem}
\begin{proof}
We first consider the case where the
input graphs do not share edges. We root the given tree in an
arbitrary leaf~$r$, which yields a directed tree~$\MT$.
We will obtain the $x$-coordinates of the vertices from a particular
post-order traversal of this directed tree. As a consequence, all
children of a vertex are placed to its left, and disjoint subtrees are
placed in disjoint $x$-intervals.
By adding dummy edges, we augment the given matching to a perfect
matching if~$n$ is even, or to a near-perfect matching if $n$ is odd.
Let~$\MM$ be the augmented matching. After the 
algorithm terminates, these dummy edges can be safely removed. 
In order to compute a \racsefe drawing of~$\MT$ and~$\MM$, we follow an 
approach that is based on the one of Cabello et al.
The difference is that (a)~rather than placing ropes in disjoint 
parallelograms, we place subtrees into disjoint axis-parallel rectangles, and
(b)~the assignment of matching edges to the two groups is much simpler.

\begin{figure}
  \begin{subfigure}[t]{.27\textwidth}
      \centering
      \includegraphics[page=1]{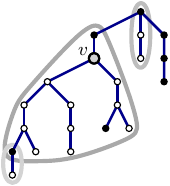}
      \caption{The placed vertices (black) induce three ropes; $v$ is a splitter.}
      \label{fig:splitter-def}
  \end{subfigure}
  \hfill
  \begin{subfigure}[t]{.68\textwidth}
      \centering
      \includegraphics[page=2]{splitter}
      \caption{Vertex~$w$ has been placed at the top in the previous
        step, creating a splitter $v$ that will now be placed at the
        bottom.} 
      \label{fig:splitter-place}
  \end{subfigure}
  \caption{Definition and placement of a splitter}
  \label{fig:splitter}
\end{figure}

First, we show how to compute the $y$-coordinates of the vertices.
Each edge of~$\MM$ is drawn horizontally at a unique odd
$y$-coordinate between~$1$ and~$n-1$. The coordinates are determined
as follows.  We start by putting the matching edge containing the root
of~$\MT$ into the top group.  Within the top group, the edges are
assigned to odd rows, from top to bottom, starting from row $n-2$
if~$n$ is odd, or $n-1$ otherwise. In the bottom group, the edges are
also assigned to odd rows, but from bottom to top, starting from
row~1. In this way, a matching edge is always assigned to an odd row
between~1 and $n-1$. Thus, our drawing needs at most~$n-1$ rows.
In the following we show how to determine the group into which a
matching edge goes.  There are two cases.

If there is no splitter, the algorithm finds an unplaced vertex that
has a tree edge to a placed vertex, and adds its matching edge to the
top group.  This creates at most one splitter since one of the
placed vertices is adjacent to a vertex that has already been placed.
Hence, whenever a new matching edge is placed, there will be at most one
splitter in the graph.

Now, assume that a splitter exists.  Call it~$v$.
For any subtree, we call the unique 
vertex with no incoming edge its \emph{local root}.
Since we start by drawing the
root of~$\MT$, every vertex lies in a rope that has its local root
placed. Let~$T_1(v),\dots,T_k(v)$ be the subtrees of~$\MT$ hanging
off~$v$; see the left drawing of Figure~\ref{fig:splitter-place}. 
Then,~$v$ is a splitter if and only if there are
two subtrees~$T_i(v)$ and~$T_j(v)$ with at least one placed vertex,
one of which has been placed in the last step. Without loss of
generality, let~$w\in T_i(v)$ be this vertex. If~$w$ was added to
the top group, we add~$v$ to the bottom group; otherwise, we add~$v$
to the top group. Thus, the vertices of the subtrees~$T_l(v)$
with~$l\neq j$ will be placed either all above or all below~$v$; 
see Figure~\ref{fig:splitter-place}.

Next, we describe how to determine the $x$-coordinates of the
vertices.  We proceed inductively.  For each non-leaf vertex, we
compute an order of the subtrees that hang off this vertex.  This
order determines the disjoint $x$-intervals into which we place the
subtrees.
Let~$v$ be a vertex that is placed in an induction step. We
traverse the path from~$v$ to the root of the rope it lies on. For
every vertex~$u$ on this path, we determine the order of the
subtrees of its children as follows. Let~$T_1(u),\ldots,T_\ell(u)$ be
the subtrees hanging off~$u$, assuming~$v\in T_1(u)$.
If~$v$ is the only placed successor of vertex~$u$, then we assign
the order~$x(T_1(u))<\ldots<x(T_\ell(u))$ to the $x$-intervals of
the subtrees; otherwise, an order has already been determined. When
the algorithm is done, we know the order of the subtrees for every
vertex on this path.  In particular, we know the $x$-interval of the
subtree rooted in~$v$. We assign the largest $x$-coordinate of this 
interval to~$v$.

Now, we show how to draw the edges. Let~$(u,v)$ be a directed edge
of~$\MT$. We draw~$(u,v)$ with a slanted segment of~$y$-length~1
at vertex~$u$ and a vertical segment at vertex~$v$.
The bend of edge lies at~$(x(v),y(v)+\sgn(y(u)-y(v)))$.
(Note that, if~$(u,v)\in\MM$, then the bend lies at~$v$, that is, the edge
is drawn horizontally.)
Since we draw the edges of the matching horizontally without bends
and since the slanted segments are drawn between two consecutive
horizontal grid lines, there can only be crossings between vertical
segments of the tree and horizontal segments of the matching. Thus,
all crossings between the tree and the matching are at right
angles.

It remains to show that the drawing of the tree itself is planar.
Since the edges of the tree are drawn with a slanted segment and a vertical
segment, each crossing must involve a slanted segment. Let~$v$ be a vertex of
the tree. We will show that the slanted segments of the edges leaving~$v$ do
not induce a crossing. Since the subtree rooted in~$v$ is assigned an
$x$-interval that contains only vertices of the
subtree, crossings can only occur with edges of this subtree.
Recall that~$v$ lies on the right border of this $x$-interval, 
so all edges leaving~$v$ are directed to the left.
Consider the step of the algorithm in which~$v$ is placed.

First, assume that~$v$ is not a splitter. 
If no successors of~$v$ have been placed so far, they will be placed
all above or all below~$v$.  Therefore, the slanted segments of the
edges leaving~$v$ won't induce any crossing.
Otherwise, let~$T_1(v),\ldots,T_k(v)$ be the subtrees hanging off~$v$. 
By construction, all placed successors of~$v$ are located
in the same subtree~$T_1(v)$, and~$T_1(v)$ is placed to the left
of the other subtrees hanging off~$v$. Thus, no edge
leaving~$v$ is drawn inside the $x$-interval assigned
to~$T_1(v)$.  The vertices in the other subtrees will be placed all
above or all below~$v$.  Therefore, the slanted segments of the edges
leaving~$v$ again won't induce any crossing.

Second, assume that~$v$ is a splitter. Then, there is a vertex~$w$
that was placed in the previous step and lies in the same rope
as~$v$; see Figure~\ref{fig:splitter-place}. 
Further, all placed successors of~$v$ except~$w$ are located
in the same subtree~$T_1(v)$, and~$T_1(v)$ is placed to the left
of the other subtrees hanging off~$v$.
Hence, no edge leaving~$v$ is drawn inside the $x$-interval 
assigned to~$T_1(v)$. Recall that~$v$ is placed in the group
opposite of~$w$. Thus,~$w$ and all unplaced successors of~$v$
(including all vertices in $T_2(v),\dots,T_k(v)$), lie
all above or all below~$v$.  Therefore, the slanted segments of the edges
leaving~$v$ do not induce any crossing.
This concludes the proof of planarity.

Note that our algorithm supports the \sefe model.  
Edges that lie both in~$\MM$ and~$\MT$ are drawn the same way;
see the edge incident to vertex~$s$ in Figure~\ref{fig:treematching}.

Finally, we prove the area and running time bounds. 
Recall that each edge of~$\MM$ is drawn horizontally at a unique odd
$y$-coordinate between~$1$ and~$n-1$. In every column, we place
exactly one vertex, so the drawing needs~$n$ columns. As for the
running time, the algorithm to place the vertices clearly requires
only constant time per vertex, except when we 
traverse the tree upwards to determine the $x$-coordinate of a new vertex.
As soon as we hit a vertex whose $x$-coordinate has already been determined,
we are done, assuming that we have precomputed the sizes of all subtrees
(which can be done in linear total time). During the traversal, we
fix the $x$-coordinates of all vertices that we meet.
Hence, in order to determine the $x$-coordinates of all vertices, 
we traverse every edge of~$\MT$ at most once. Thus, the running time of
the algorithm is linear.
\end{proof}

\section{\racsefe Drawings with Two Bends per Edge}
\label{sec:twobends}

In this section, we study slightly more complex classes of planar
graphs, and show how to efficiently construct \racsefe drawings with
two bends per edge, in quadratic area. In particular, we prove that
a wheel and a matching on a common set of $n$ vertices admit a
\racsefe drawing on an integer grid of size $(1.5n-1) \times (n+2)$
with two bends per wheel edge and no bends on matching edges;
see~Theorem~\ref{thm:wheelmatch}. 
An outerpath (that is, an outerplanar graph whose weak dual is a
path) and a matching admit a \racsefe drawing with two bends per
outerpath edge and one bend per matching edge.  They also need a
slightly larger grid, namely one of size $(3n-2) \times (3n-2)$; see
Theorem~\ref{thm:outpmatch}.

\begin{theorem}
    A wheel and a matching on a common set of $n$ vertices admit a
    \racsefe drawing on an integer grid of size $(1.5n-1) \times (n+2)$
    with two bends per wheel edge and no bends on matching edges, respectively.
    The drawing can be computed in $O(n)$ time.
    \label{thm:wheelmatch}
\end{theorem}
\begin{proof}
Let $\MW=(V,E_\MW)$ be the given wheel and $\MM=(V,E_\MM)$
the given matching.
A wheel can be decomposed into a cycle, called its \emph{rim}, a
center vertex, called its \emph{hub}, and a set of edges that
connect the hub to the rim, called its \emph{spikes}. First, we
consider the simpler case according to which $\MW$ and $\MM$ do not
share edges (clearly, in this case the hub of the wheel cannot be
incident to a matching edge). Let $V = \{v_1, v_2, \ldots, v_n\}$,
such that~$v_1$ is the hub of $\MW$ and $\MC=\langle v_2, v_3,
\ldots, v_{n}, v_2\rangle$ is the rim of~$\MW$ in this order.  Thus,
$E_\MW = \{(v_i,v_{i+1})\mid i = 1,\ldots,n-1\} \cup \{(v_n, v_2)\}
\cup \{(v_1,v_i)\mid i = 2,\ldots,n\}$.
Let~$\MM'=(V,E_{\MM'})$ be the matching~$\MM$ without the edge
incident to~$v_1$.

\begin{figure}[tb]
	\centering
	\includegraphics{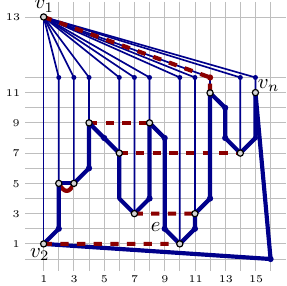}
	\caption{A \racsefe drawing of a wheel (solid; its rim is drawn in bold) and a 
		matching (dashed)}
	\label{fig:wheelmatching} 
\end{figure}

We first compute the $x$-coordinates of the vertices.  We do this such
that $\MC-\{(v_n,v_2)\}$ is $x$-monotone;
see Figure~\ref{fig:wheelmatching}. More
precisely, for $i\in\{2,\ldots, n\}$, we set $x(v_i)=2i-3$.
The $y$-coordinates of the vertices are computed based on the
matching $\MM'$ as follows. Let~$E_{\MM'}=\{e_1,\ldots,e_k\}$ be
the matching edges, indexed such that~$v_2$ is incident to~$e_1$.
For~$j\in\{1,\ldots,k\}$, we assign the $y$-coordinate~$2j-1$ to the
endpoints of~$e_j$. Next, we assign the $y$-coordinate~$2k+1$ to
the vertices incident to the rim without a matching edge in~$\MM'$.
Finally, the hub~$v_1$ of~$\MW$ is located at point~$(1,2k+3)$.

It remains to place the bend of each edge~$e\in E_\MW$;
the edges in~$\MM'$ are drawn without bends.
First, let~$e=(v_1,v_i)$,
$i\in\{3,\ldots,n\}$ be a spike. Then, we place the bend
at~$(x(v_i),2k+2)$. Since both~$v_1$ and~$v_2$ are located in
column~1, we can save the bend of the spike $(v_1,v_2)$. Second,
let~$e=(v_i,v_{i+1})$, $i\in\{2,\ldots, n-1\}$ be an edge of the
rim~$\MC$. 
We place the bend according to the following case distinction.
\begin{enumerate}[(i)]
\setlength{\itemsep}{0mm}
\item If~$y(v_{i+1})>y(v_i)$, we place the bend
at~$(x(v_{i+1}),y(v_{i})+1)$.
\item If~$y(v_{i-1})>y(v_i)>y(v_{i+1})$, we
place the bend at~$(x(v_{i+1}),y(v_{i})-1)$.
\item If~$y(v_i)>\max\{y(v_{i-1}),y(v_{i+1})\}$, by (i) the bottom port
  at~$v_i$ is already used; see the edge~$e$ in
  Figure~\ref{fig:wheelmatching}.  Thus, we draw~$e$ with two bends;
  at~$(x(v_i)+1,y(v_{i})-1)$ and~$(x(v_i)+1,y(v_{i+1})+1)$.
\end{enumerate}
Now, let~$e=(v_n,v_2)$ be the remaining edge of the rim.
We place its bend at~$(2n-2,0)$.

Our approach ensures that $\MC-\{(v_n,v_2)\}$ is drawn
$x$-monotone, hence planar. The last edge $(v_n,v_2)$ of~$\MC$ is
the only edge drawn outside of the bounding box that contains 
all vertices.  Therefore, also the last edge is crossing-free.
The spikes are not involved in crossings
with the rim, as they are outside of the bounding box containing
the rim edges. Hence, the drawing of $\MW$ is planar. All
edges of $\MM'$ are drawn as horizontal, non-overlapping line
segments, so $\MM'$ is drawn planar as well. The slanted
segments of $\MW-{(v_n,v_2)}$ are of $y$-length~1, so they cannot
be crossed by the edges of $\MM'$. As the edge~$(v_n,v_2)$ is not
involved in crossings, it follows that all crossings between~$\MW$
and~$\MM'$ form right angles.

Finally, we have to insert the matching edge~$(v_1,v_i)$ 
incident to the hub.  Note that this edge also exists in \MW as a
spike. Since~$v_i$ is not incident to a
matching edge in~$\MM'$, it is placed above all matching edges.
Therefore, the copy of $(v_1,v_i)$ in~\MW does not cross a matching edge, 
and we can use the same layout for the copy of $(v_1,v_i)$ in~\MM.

We now show that our algorithm supports the \sefe model. 
Suppose that there is an edge~$e=(v,v')$
that belongs to both the rim~\MC and the
matching~\MM. If~$e$ is the closing edge of the rim, then it is
drawn planar in~$\MC$. Thus, this drawing can be used for both
graphs. Otherwise, our algorithm places~$v$ and~$v'$ at consecutive
$x$-coordinates and at the same $y$-coordinate. Then, we can
draw~$e$ as a horizontal edge of length~1 in both graphs, and such
an edge cannot be crossed.

We now prove the area bound of the drawing algorithm. To that
end, we remove all columns that contain neither a vertex, nor a
bend. First, we count the rows used. Since we remove the matching
edge incident to~$v_1$, the matching~$\MM'$ has~$k\le n/2-1$
matching edges. We place the bottommost vertex in row~$1$ and the
topmost vertex (that is, vertex~$v_1$) in row~$2k+3$. We add one
extra bend in row~$0$ for the edge~$(v_n,v_2)$. Thus, our drawing
uses $2k+3+1\le n+2$ rows. Next, we count the columns used. The
vertices~$v_2,\dots,v_n$ are each placed in their own column. Every
spike has exactly one bend in the column of a vertex. An
edge~$(v_i,v_{i+1})$ of rim $\MW$ has exactly one bend in a vertex
column, except for the case that~$y(v_i)>y(v_{i-1}),y(v_{i+1})$.
In this case it needs an extra bend between~$v_i$ and~$v_{i+1}$, $i =
1,\ldots,n-1$. Clearly, there can be at most~$n/2-1$ vertices
satisfying this condition. Since the edge~$(v_n,v_2)$ uses an extra
column to the right of~$v_n$, our drawing uses
$(n-1)+(n/2-1)+1=1.5n-1$ columns.
\end{proof}

\begin{figure}[tb]
  \begin{subfigure}[t]{.34\textwidth}
    \centering
    \includegraphics[page=1]{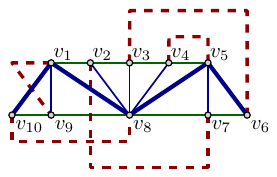}
    \caption{A non-planar drawing}
    \label{fig:outerpathmatching-1}
  \end{subfigure}
	\hfill
  \begin{subfigure}[t]{.6\textwidth}
    \centering
    \includegraphics[page=2]{outerpathmatching}
    \caption{A \racsefe drawing}
    \label{fig:outerpathmatching-2}
  \end{subfigure}
  \caption{Two drawings of the same outerpath and matching. In both figures, the 
		outerpath is drawn solid, the upper and the lower path are drawn in bold, the 
		spine of the spanning caterpillar is drawn extra bold and the matching is 
		drawn dashed.}
  \label{fig:outerpathmatching}
\end{figure}

\begin{theorem}\label{thm:outpmatch}
  An outerpath and a matching on a common set of $n$ vertices admit a \racsefe drawing
  on an integer grid of size $(3n-2) \times (2n-1)$
  with two bends per outerpath edge and one bend per matching edge.
  The drawing can be computed in $O(n\log n)$ time.
\end{theorem}
\begin{proof}
Let~$\MZ=(V,E_\MZ)$ be the given outerpath and $\MM=(V,E_\MM)$
the given matching. Recall that an outerpath is a biconnected outerplanar
graph whose weak dual is a path of length at least two; see
Figure~\ref{fig:outerpathmatching-1}. Furthermore, assume initially
that $\MZ$ and $\MM$ do not share edges. Let $V =
\{v_1,v_2,\ldots,v_n\}$, indexed such that $\langle v_1, v_2,
\ldots, v_n, v_1\rangle$ is the outer face of~$\MZ$.

We start by augmenting~$\MZ$ to a maximal
outerpath~$\MZ'=(V,E_{\MZ'})$ by triangulating its bounded faces. As
$\MZ'$ is internally-triangulated, it contains exactly two vertices
of degree two, each of which lies on a face that is 
an endpoint of the dual path.  We assume, without loss of generality,
that~$\deg(v_n)=\deg(v_j)=2$ for some~$j$ with~$1<j<n$; see~$v_{10}$
and~$v_6$ in Figure~\ref{fig:outerpathmatching-1}. We call the path
$\MP_\mathrm{u} = (V_\mathrm{u},E_\mathrm{u}) = \langle v_1, v_2, \ldots, v_{j-1} \rangle$ the
\emph{upper path} of~$\MZ'$, and~$\MP_\mathrm{l} =
(V_\mathrm{l},E_\mathrm{l}) = \langle v_j, v_{j+1}, \ldots, v_n \rangle$ the
\emph{lower path} of~$\MZ'$. Observe that $V=V_\mathrm{u} \cup V_\mathrm{l}$.
If we remove~$E_\mathrm{u} \cup E_\mathrm{l}$ from~$E_{\MZ'}$, then the
resulting graph is a caterpillar~$\MC$ that spans~$V$ and whose
spine alternates between vertices of~$V_\mathrm{u}$
and~$V_\mathrm{l}$.

We first compute the left-to-right order of the vertices of
caterpillar~$\MC=(V_\MC,E_\MC)$ as described by the algorithm
supporting Theorem~\ref{thm:cater}. Then, the $x$-coordinate of the
$i$\textsuperscript{th}
vertex in this order is~$3i-2$, $i=1,2,\ldots,n$; see
Figure~\ref{fig:outerpathmatching-2}.

In order to compute the $y$-coordinates of the vertices, we first
partition~$\MM$ into three matchings~$\MM_\mathrm{ll}=(V_\mathrm{ll},
E_\mathrm{ll})$, $\MM_\mathrm{uu}=(V_\mathrm{uu},E_\mathrm{uu})$ and $\MM_\mathrm{ul}=(V_\mathrm{ul},
E_\mathrm{ul})$ as follows. Let~$(v,v')\in E_\MM$. Then,
\begin{enumerate}[(i)]
    \item $(v,v')\in E_\mathrm{ll}$ if~$v,v'\in V_\mathrm{l}$,
    \item $(v,v')\in E_\mathrm{uu}$ if~$v,v'\in V_\mathrm{u}$,
    \item $(v,v')\in E_\mathrm{ul}$ if~$v\in V_\mathrm{u}$ and~$v'\in V_\mathrm{l}$.
\end{enumerate}
Since $V=V_\mathrm{u} \cup V_\mathrm{l}$, it holds that~$E_\MM = E_\mathrm{ll} \cup
E_\mathrm{uu}\cup E_\mathrm{ul}$. In the resulting layout, the edges
in~$E_\mathrm{ll}$ will be drawn below the edges in~$E_\mathrm{ul}$,
which in turn will be drawn below the ones of $E_\mathrm{uu}$; see
Figure~\ref{fig:outerpathmatching-2}. Thus, they will not cross each
other.

Let~$m_\mathrm{ll}=|E_\mathrm{ll}|$
and~$E_\mathrm{ll}=\{e_1,\ldots,e_{m_\mathrm{ll}}\}$. We draw the
edges from bottom to top, starting from row~1. For
$i=1,\ldots,m_\mathrm{ll}$, 
let~$e_i=(u_i,u_i')$ with~$x(u_i)<x(u_i')$. We set~$y(u_i)=4i-1$ and
$y(u_i')=4i-3$. Edge~$e_i$ is drawn with a bend
at~$(x(u_i),y(u_i'))$. Our approach ensures that there are no
crossings between edges of $\MM_\mathrm{ll}$, as they are drawn in
different horizontal strips of the drawing. Similarly, we draw the
edges in~$E_\mathrm{uu}$ from top to bottom, starting from row~$2n-1$.
Let~$m_\mathrm{uu}=|E_\mathrm{uu}|$. By construction, the topmost vertex
of~$V_\mathrm{ll}$ is drawn in the row~$4m_\mathrm{ll}-1$ and the
bottommost vertex of~$V_\mathrm{uu}$ is drawn in the row~$2n+1-4m_\mathrm{uu}$.
The vertices of~$V_\mathrm{ul}$ will
be drawn between these rows; see Figure~\ref{fig:outerpathmatching-2}.

In order to draw the edges in~$E_\mathrm{ul}$, we process the vertices
of the set~$V_\mathrm{ul}$ from left to right and assign $y$-coordinates
to both endpoints of the incident matching edge.
Let~$m_\mathrm{ul}=|E_\mathrm{ul}|$
and~$E_\mathrm{ul}=\{\eps_1,\ldots,\eps_{m_\mathrm{ul}}\}$. For~$k\in\{1,\ldots,\mu\}$,
let~$\eps_k=(w_k,w_k')$ and assume without loss of generality
that~$x(w_k)<x(w_k')$ and~$x(w_1)<\ldots<x(w_\mu)$. We place the
vertices in~$V_\mathrm{l}$ from bottom to top and the vertices
in~$V_\mathrm{u}\cap V_\mathrm{ul}$ from top to bottom. If~$w_k\in V_\mathrm{u}$, we
assign the $y$-coordinate $4m_\mathrm{ll}-1+3k$ to~$w_k$ and the
$y$-coordinate $2n+1-4m_\mathrm{uu}$ to~$w_k'$; if~$w_k\in V_\mathrm{l}$, we
switch the $y$-coordinates. Edge~$\eps_k$ is drawn with a bend
at~$(x(w_k),y(w_k'))$. Further, every edge $\eps_l\in E_\mathrm{ul}$
with~$l>k$ has its endpoints to the right of~$w_k$ and in the
horizontal strip defined by the lines~$y=y(w_k)$ and~$y=y(w_k')$.
Hence, it will not be involved in crossings with~$(w_k,w_k')$. This
guarantees that the drawing of $\MM$ is planar.

It remains to determine, for each edge~$e=(v,v')\in\MZ'$, where to
place its bend.
Without loss of generality, let~$e$ be directed from its left
endpoint, say~$v$, to its right endpoint, say~$v'$. First, assume
that~$e\in E_\mathrm{u}\cup E_\mathrm{l}$ belongs to the outercycle. Then, we place
its bends at~$(x(v')-2,y(v))$
and~$(x(v')-2,y(v')-\sgn(y(v')-y(v)))$. Second, assume that~$e\in
E_\MC$ belongs to the inner caterpillar. Then, we place its bends
at $(x(v')-1,y(v)+\sgn(y(v)-y(v')))$
and $(x(v')-1,y(v')-\sgn(y(v')-y(v)))$.

Since~$\MP_\mathrm{u}$ and~$\MP_\mathrm{l}$ are drawn $x$-monotone, both are drawn
planar. Following similar arguments as in the proof of
Theorem~\ref{thm:cater}, we can show that~$\MC$ is drawn planar as
well. Since~$\MC$ is drawn between~$\MP_\mathrm{u}$ and~$\MP_\mathrm{l}$, it
follows that~$\MZ'$ is drawn planar, as desired. It now remains to
prove that all (potential) crossings between~$\MZ'$ and~$\MM$ only
involve rectilinear edge segments of~$\MZ'$, as~$\MM$ consists
exclusively of rectilinear segments. As all slanted segments
of~$\MZ'$ are of $y$-length~1, no horizontal segment of~$\MM$ can
cross them. The same holds for vertical segments of $\MM_\mathrm{uu} \cup
\MM_\mathrm{ll}$, as they are drawn above and below $\MP_\mathrm{l}$ and
$\MP_\mathrm{u}$, respectively. The only possible non-rectilinear crossings
are between a vertical segment of a matching edge~$(u,v)\in
E_\mathrm{ul}$ and a long slanted segment of~$\MC$ incident to a spine
vertex~$w$. This crossing can only occur if~$w$ lies to the left of
the vertical segment of~$(u,v)$. By construction,~$w$ is never drawn
between~$u$ and~$v$, with respect to the $y$-coordinate. Thus, such
crossings cannot occur, which implies all crossings between $\MZ'$
and $\MM$ form right angles.

We now show that our algorithm supports the \sefe model.  
Assume that~\MM and~\MZ share edges, and let~$e=(v,v')$
be an edge that belongs to both input graphs.
If~$e\in E_\mathrm{l}$, then also~$e\in E_\mathrm{ll}$. Our algorithm 
places~$v$ and~$v'$ at consecutive
$y$-coordinates and determines their $x$-coordinates such that no other vertex 
of~$V_\mathrm{l}$ lies between them. Thus, edge~$e$ is drawn in the matching 
as a vertical segment of length~2 and a horizontal segment that has no vertex 
below it. Hence, the drawing of~$e$ in the matching is planar and can be used
for both graphs. The case~$e\in E_\mathrm{u}$ is analogous.
 
If~$e\in E_\MC$, then also~$e\in E_\mathrm{ul}$. In this case, we observe 
that the topological routing of~$e$ is the same in both graphs with an offset
of one unit to avoid overlaps. Thus, every other edge either crosses both
or none of the drawings of~$e$. Since each drawing of~$e$ forbids crossings 
by edges of one of the input graphs, actually none of the two drawings
can have a crossing.  Hence, we can draw~$e$ the same way in both graphs.

By the choice of the coordinates, the area requirement of our
algorithm is $(3n-2)\times(2n-1)$. Since we have to sort the edges
in~$E_\mathrm{ul}$ by the $x$-coordinates of the incident vertices, our
algorithm runs in~$O(n\log n)$ time. To complete the proof of this
theorem, observe that the extra edges that we introduced when
augmenting~$\MZ$ to~$\MZ'$ can be safely removed from the
constructed layout, affecting neither the crossing angles nor
the area of the layout.
\end{proof}

\section{Conclusions and Open Problems}

In this paper, we have studied \racsefe and \racsim drawings with few
bends per edge.  We have proven that two planar graphs always admit a
\racsim drawing with at most six bends per edge.  For more restricted
classes of graphs, we have reduced the number of bends per edge.  Some
of these specialized results also hold for the stronger \racsefe
model.  All drawings of our algorithms fit it quadratic area.

Our results raise several questions that remain open.  First of all,
can we strengthen any of our results from \racsim to \racsefe (see
Table~\ref{table:res}); for example, for a pair of general planar
graphs?  For the classes
of graphs that we have presented in this paper, can we reduce the
number of bends per edge or can we show lower bounds?  Are there other
graph classes that admit \racsim drawings with few bends?  Can we
reduce the number of bends per edge by relaxing the strict constraint
that edge intersections are at right angles and instead ask for
drawings that have close to optimal crossing resolution?  What about
the computational complexity of the general problem, that is, given
two or more planar graphs on the same set of vertices and a
non-negative integer~$k$, can we decide efficiently whether there is a
\racsim drawing in which each graph is drawn with at most~$k$ bends
per edge and the crossings are all at right angles?  This seems
unlikely.  Finally, is it possible to achieve sub-quadratic area for
\racsim drawings of subclasses of planar graphs when accepting more
bends per edge?

\newpage

\bibliographystyle{abbrvurl}
\bibliography{abbrv,racsim}

\end{document}